\documentstyle[preprint,aps,eqsecnum]{revtex}

\tightenlines

\def\m@thcombine#1#2{%
  \setbox0=\hbox{$#1$}
  \setbox1=\hbox{$#2$} 
  \ifdim\wd0>\wd1
    \setbox0=\hbox to\wd1{\hss\box0\hss}
  \else
    \setbox1=\hbox to\wd0{\hss\box1\hss}
  \fi
  \mathop{\vcenter{
    \offinterlineskip\box0\box1}}}
\def\lesim{\m@thcombine<\sim}
\def\gesim{\m@thcombine>\sim}
\begin{document}


\draft
\title{ CHIRAL TOPOLOGY IN THE ZERO MODES ENHANCEMENT QUANTUM MODEL OF THE      
        QCD VACUUM }

\author{V. Gogohia\footnote{On leave of abcence from RMKI, KFKI, Budapest, Hungary} }

\address{Reserach Center for Nuclear Physics (RCNP), Osaka University \\
          Mihogaoka 10-1, Ibaraki, Osaka 567-0047, Japan }

\maketitle

\begin{abstract}
Using the effective potential approach for composite operators we have formulated the quantum model of the QCD vacuum. It is
based on the existence and importance of the nonperturbative $q^{-4}$ quantum, 
topologically nontrivial excitations of the gluon field configurations there.  
As a result of this the QCD vacuum is found stable and it has a stationary state. 
The value of the scale responsible for the nonperturbative dynamics is taken from the bounds for the pion decay constant in the chiral limit. We have obtained
good agreement with the phenomenological values of the topological susceptibility, the mass of $\eta'$ meson, the gluon condensate. An excellent agreement with phenomenological values of the above mentioned quantities is achived by simply summing up our contributions and instanton-induced contributions due to Negele et al to the vacuum energy density.  
\end{abstract}


\vfill

\eject

\section{Introduction }

The nonperturbative QCD vacuum has very rich dynamical
and topological structure. It is a very complicated medium and its dynamical and topological complexity means that its structure can be organized at various levels (classical, quantum, etc) and it can contain many different components and ingredients which contribute to the vacuum energy density, the one of main characteristics of the QCD ground state.  
Many models of the QCD vacuum involve some extra classical colored fiel 
configurations such as randomly oriented domains of a constant color magnetic 
fields [1], background gauge fields, averaged over spin and color [2], stochastic colored background fields [3], etc (see Ref. [4] and references therein).   
The most elaborated random instanton liquid model (RILM) of the QCD vacuum [5-7] is based on the existence of the topologically nontrivial instanton-type fluctuations of gluon fields, which are solutions to the classical equations of motion in Euclidean space [8]. 

On the other hand, today there are no doubts left that the dynamical mechanisms
of the important nonperturbative quantum phenomena such as quark confinement and dynamical (or equivalently spontaneous) chiral symmetry breaking (DCSB) are closely related to the complicated topologically nontrivial structure of 
the QCD vacuum [9-11]. For this reason, any correct nonperturbative model of
quark confinement and DCSB necessary turns out to be a model of the true QCD vacuum and the other way around. Our model of
the true QCD ground state is based on the existence and importance of the nonperturbative $q^{-4}$ behaviour of the full gluon propagator at small $q^2$, which is a manifestation of completely $quantum$ excitations of gluon field configurations there. It describes the zero modes enhancement (ZME) effect in QCD at large distances [12,13] (for additional references see Ref. [14]). These excitations also are topologically nontrivial in
comparison with the free gluon structure, $q^{-2}$. In this context, let us note that the attractive classical model of the QCD vacuum as a condensation
of the color-magnetic monopoles (QCD vacuum is a chromomagnetic superconductor)
proposed by Nambu, Mandelstam and 't Hooft and developed by Nair and Rosenzweig
(see Ref. [15] and references therein) as well as the classical mechanism of   
the confining medium [16] and an effective theory for the QCD vacuum proposed  
in [17], also invokes $q^{-4}$ behaviour of the gluon fields in the IR. 
Let us underline that without $q^{-4}$ component in
the decomposition of the full gluon propagator it is impossible to obtain the  
area law for static quarks (indicative of confinement)
within the Wilson loop approach [18]. This behaviour of the full gluon propagator in the IR is also required to derive the heavy quark potential within the recently proposed exact renormalization group approach [19].    

Our approach to nonperturbative QCD is based on solutions to
the quark and ghost Schwinger-Dyson (SD) quantum equations of motion which should be complemented by the investigation of the corresponding Slavnov-Taylor (ST) identities [14,20,21]. Such sigular behaviour of the full gluon propagator in
the infrared (IR) region requires the introduction of a small IR regulation parameter $\epsilon$, in order to define the initial equations in the IR by the dimensional regularization method [22] within the distribution theory [23]. This 
yeilds the regularization expansion for the above mentioned strong IR singularity as follows (four dimensional Euclidean space) [14,20,21,23]

\begin{equation}
(q^2)^{-2 + \epsilon} = {\pi^2 \over \epsilon} \delta^4 (q) + finite \ terms,
\qquad \epsilon \rightarrow 0^+
\end{equation}
and the terms of order $\epsilon$ are not shown here for simplicity. Because of
this, the quark propagator and other Green's function become dependent, in general, on this IR regulation parameter $\epsilon$, which is to be set to zero at 
the end of computations, $\epsilon \rightarrow 0^+$. For the sake of brevity, this dependence is always understood but not indicated explicitly.              

 There are only two different types of
behaviour of the quark propagator with respect to $\epsilon$ in
the $\epsilon \rightarrow 0^+$ limit.
 If the quark propagator does not depend
on the $\epsilon$ - parameter in the $\epsilon \rightarrow 0^+$
 limit then one obtains the IR regularized
(from the very beginning) quark propagator. In this
case quark confinement is understood as the disappearance of the
quark propagator pole on the real axis at the point $p^2 = m^2$,
where $m$ is the quark mass. Such an understanding (interpretation) of quark confinement comes, apparently from Preparata's massive quark model (MQM) [24] in 
which quarks were approximated by entire functions. A quark propagator may or may not be an entire function, but in any case the pole of the first order
(like the electron Green's function has in QED) disappears (see Ref. [14] and references therein). However, $the \ absence \ of \ the \ pole-type \ singularities$ in the quark Green's functions is only the $first \ necessary$ condition of the quark confinement at the fundamental (microscopic) quark-gluon level. At hadron (macroscopic) level there exists the $second \ sufficient$ condition, 
namely the corresponding Bethe-Salpeter (BS) equation for the bound-states should have the $discrete \ spectrum \ only$ [25] in order to prevent quarks to appear in asymptotically free states. At nonzero temperatures and densities, for example in quark-gluon plasma (QGP), the bound-states will be dissolved, but, nevertheless the first necessary condition still remains valid, of course. Thus in general case confinement criterion consists of the two above formulated parts. This definition generalises the linier rising potential between heavy quarks 
since it is relevant not only for light quarks but for heavy quarks as well.   
On the other hand, a quark propagator can vanish after the removal ($\epsilon \rightarrow 0^+$) of the IR regulation parameter $\epsilon$. A  vanishing quark 
propagator is also a direct manifestation of quark confinement (see again Ref. [14] and references therein). Such understanding of quark confinement comes,
apparently, from two-dimensional QCD with $N_c$ large limit [25]. 

We develop a method for the extraction of the IR finite (regularized) Green's functions in QCD. The IR finiteness of the Green's functions means that they exist as $\epsilon \rightarrow 0^+$. For this purpose, we have worked out a renormalization program in order to cancel all IR nonperturbative divergences, which makes it possible to explicitly show that all Green's functions
are IR multiplicative renormalizable (MR). On the other hand this leads to a closed set of equations in the quark sector [14,20,21]. We first approximate the 
exact quark SD equation by its deep IR (confinement) piece assuming that precisely this term is mainly responsible for nonperturbative effects in
QCD in particular quark confinement. Introducing the appropriate dimensionless 
variables, the quark SD equation for the quark propagator $iS(p) = \hat p A(p^2) - B(p^2)$, becomes

\begin{equation}
xA' = - (2+x)A - 1 - m_0 B,                
\end{equation}
and

\begin{equation}
2BB' = - 3 A^2 + 2(m_0A-B)B,               
\end{equation}
where $A,B \equiv A(x),B(x)$ and the prime denotes the derivative with respect
to the Euclidean dimensionless momentum variable, $x= p^2/ \bar \mu^2$, where  
$\bar \mu^2$ is the characteristic scale coming from $q^{-4}$ behaviour of the full gluon propagator in the IR. In the chiral limit     
$m_0=0$, it can be solved exactly and solutions are 

\begin{equation}
A(x) = x^{-2}( 1 - x - e^{-x})                                                
\end{equation}
and 

\begin{equation}
B^2(x_0, x) = 3e^{-2x} \int^{x_0}_x dx'e^{2x'} A^2(x'),                        
\end{equation}
where $x_0=p_0^2/ \bar \mu^2$ is an arbitrary constant of integration. It is easy to see that obtained solution for the quark propagator is regular at zero point, has no pole-type singularities (indicative of confinement) and corresponds
to dynamical breakdown of chiral symmetry (quark mass generation) since the chiral symmetry violating solution ($m_0=0, A(x) \ne 0, B(x) \ne 0$) is only allowed (chiral symmetry preserving solution ($m_0=B(x)=0, A(x) \ne 0$) is forbidden). It is also nonperturbative ( it can not be expanded in powers of the coupling constant) and the function $A(x)$ automatically approaches the free propagator at infinity. In order to reproduce a
correct behaviour at infinity ($x \rightarrow \infty$) of the dynamically generated quark mass function, it is necessary to put $x_0=\infty$ in
(1.5) from the very beginning, so it identically vanishes in the limit $x \rightarrow \infty$ in accordance with the vanishing current quark mass in the chiral limit. In this case the solution (1.5) cannot be accepted at the origin $x=0$. Hence, we have to keep the constant of integration $x_0$ in (1.5) arbitrary but
finite in order to obtain a regular and nontrivial solution for the deep IR region, $x_0 \ge x$. It exhibits an algebraic branch points at $x=x_0$ and at infinity, which are caused by the inevitable ghost contributions in the covarint gauge. However, our solution does not explicitly depend on ghost degrees of freedom as well as on gauge choice.

\section{The vacuum energy density }

   As it was mentioned in Introduction any correct nonperturbative model of
quark confinement and DCSB necessarily becomes a model of the QCD ground state,
i. e. its nonperturbative vacuum.
The effective potential approach for composite operators [26,27]
allows us to investigate the QCD vacuum, since in the
absence of external sources the effective potential is nothing
but the vacuum energy density, one of the main characteristics of the nonperturbative vacuum. Let us start with the effective potential to leading order (log-loop level) in the quark sector [26] 

\begin{equation}
V(S) =- i \int {d^np \over {(2\pi)^n}} Tr \left\{
 \ln (S_0^{-1}S) - (S_0^{-1}S) + 1 \right\},
\end{equation}
where $S(p)$ and $S_0(p)$ are the full and free quark propagators,
respectively. Here and everywhere below in this section the trace over space-time and color group indices is understood. Let us note that the effective potential (2.1) is normalized as follows  $V(S_0)= 0$.

In order to evaluate (2.1) we use the well-known expression,

\begin{equation}
 Tr \ln (S_0^{-1}S) = 3 \times \ln det (S_0^{-1}S) =
 3 \times 2 \ln p^2 \left[ p^2 A^2(-p^2) - B^2(-p^2) \right],
\end{equation}
where $p^2 A^2(-p^2) - B^2(-p^2)  = \sqrt {det[-iS(p)]}$.
The factor 3 comes from the trace over quark color
indices. Going over to Euclidean space ($d^4p \rightarrow i d^4p, \quad p^2 \rightarrow -p^2$ ), in terms of dimensionless variables and functions (1.2-1.5), 
we finally obtain after some algebra ($n=4$),
  
\begin{equation}
\epsilon_q = V(A, B) =  {3 \over 8 \pi^2} p_0^4 x_0^{-2}
\int \limits_0^{x_0} dx\, x\, \left\{ \ln (x [x A^2(x) +
B^2(x_0, x)])  + 2 x A(x) + 2 \right\},
\end{equation}                                                                 
where we need to identify the ultraviolet (UV) cut-off with the constant of integration $x_0$ in order to be guaranteed that the unphysical singularities (algebraic branch points, mentioned above in section
1) in the $B^2(x_0, x)$ function (1.5) will not affect the effective potential
which should be always real in order to avoid the vacuum instability [28].

The constant of integration  $x_0$ should be related to the confinement scale at which nonperturbative effects become essential. For this reason,  within our 
approach to QCD at large distances in order to obtain numerical values
of any physical quantity, e.g. the pion decay
constant (see below and Ref. [29]), the integration over the whole range
$\left[ 0, \  \infty \right] $ reduces to the integration over the nonperturbative region $ \left[ 0, \  x_0 \right], $  which determines the range of
validity of the deep IR asymptotics (1.2) of the full gluon
propagator and consequently the range of validity of the
corresponding solutions (1.4) and (1.5) for the confinement piece of
the full quark propagator. Playing a role of the UV cut-off, the
arbitrary constant of integration $x_0$ thereby prevents the quark
propagator from having an imaginary part. This is consistent
with the idea that a confined particle should have no imaginary part [30].
We emphasize that the main contribution to the values of 
the physical quantities comes from the nonperturbative region
(large distances), whereas the contributions from the
short and intermediate distances (perturbative region), because of less singular behaviour in the IR, can only be treated as perturbative corrections. We
confirm this physically reasonable assertion numerically
in our numerical calculations.

We evaluate the nonperturbative gluon part of the effective potential, which at
the log-loop level is given by [26]

\begin{equation}
V(D) =  { i \over 2} \int {d^np \over {(2\pi)^n}}
 Tr\{ \ln (D_0^{-1}D) - (D_0^{-1}D) + 1 \}, 
\end{equation}
where $D(p)$ is the full gluon propagator and $D_0(p)$ is its  
free (perturbative) counterpart.
The effective potential is normalized as $V(D_0) = 0$, i. e. as in the quark case the perturbative vacuum is normalized to zero.
In a similar way to Eq. (2.2), we obtain

\begin{equation}
 Tr \ln (D_0^{-1}D) = 8 \times \ln det (D_0^{-1}D) =
 8 \times 4 \ln \left[ {3 \over 4 }d(-p^2) + {1 \over 4 } \right],
\end{equation}
where the factor 8 is due to the trace over the gluon colour indices and it becomes zero (in accordance with the above mentioned normalization condition) when
the full gluon form factor is replaced by its free counterpart by setting simply $d(-t^2, a) = 1$. Approximating now the full gluon form factor by its deep IR
(confinement) piece,  namely $d(-p^2)= \bar \mu^2 / (- p^2)$
and after doing some algebra in terms of new variables and parameters (1.2-1.5), we finally obtain (in Euclidean space, $n=4$) the following expression for the vacuum energy density due to nonperturbative gluon contributions, 
$\epsilon_g = V(D)$ 
    
\begin{equation}
\epsilon_g = {1 \over \pi^2} p_0^4 x_0^{-2} \times I^a_g(0, z_0),
\end{equation}
where    

\begin{equation}
I^a_g (0, z_0) = \int \limits_0^{z_0} dx\, x\, \Bigl\{  \ln x - \ln (x+ 3)     
+ {3 \over 4x} - a  \Bigr\} 
=  {9 \over 2} \ln(1 + {z_0 \over 3})
- {3 \over 4} z_0 - {1 \over 2} z^2_0 \ln(1 + {3 \over z_0})                 
-  {a \over 2} z^2_0
\end{equation}
and $a = (3/4) - 2 \ln 2 = - 0.6363$. In this equation $z_0$ is, of course, the
corresponding UV cut-off which in general differs (i. e. independent) from the 
quark cut-off $x_0$.   

 The effective potential at the log-loop level for the ghost degrees of freedom
is [26]

\begin{equation}
V(G) = - i \int {d^np \over {(2\pi)^n}}
Tr\{ \ln (G_0^{-1} G) - (G_0^{-1} G) + 1 \}, 
\end{equation}
where $ G(p)$ is the full ghost propagator and $G_0(p)$ is its  
free (perturbative) counterparts. 
The effective potential $V(G)$ is normalized as 
$V(G_0) = 0$, i.e. here like in quark and gluon cases the energy of the perturbative vacuum is set zero.
Evaluating the ghost term $\epsilon_{gh} = V(G)$ (2.8) in a very similar
way, we obtain
 
\begin{equation}
\epsilon_{gh} =  {1 \over \pi^2} p_0^4 x_0^{-2} \times I_{gh}(0, y_0),
\end{equation}
where the integral $I_{gh}(0, y_0)$ depends on the IR renormalized ghost self-energy, which remains arbitrary (unknown) within our approach. We have introduced the ghost UV cut-off $y_0$ as well.    

 In principle, we must sum up all contributions in order to obtain total vacuum
energy density. However, $\epsilon_g$ and $\epsilon_{gh}$ are divergent and therefore they depend completely on arbitrary UV cut-offs $z_0$ and $y_0$ which have nothing to do with $x_0$. Thus the sum $\epsilon_g + \epsilon_{gh}$ should be regularized in order to define finite  $\epsilon^{reg}_g$ which will depend only on $x_0$ and at the same time it should be a vanishing function of $x_0$ at
$x_0 \rightarrow \infty$ (at fixed $p_0$, see discussion below) because of the 
normalization of the perturbative vacuum to zero. For this purpose, 
we decompose the integral (2.6-2.7) into the three parts as follows $I^a_g(0, z_0) =I^a_g(0, x_0) + I^a_g(x_0, z_0)= I_g(0, x_0) - (a/2) x_0^2 + I^a_g(x_0, z_0)$.
Using the arbitrarness of the ghost term (2.9), let us 
substract the unknown integral $- (a/2) x_0^2 + I^a_g(x_0, z_0)$ from the gluon
part of the vacuum energy density by imposing  the following condition $\Delta =-(a/2) x_0^2 + I^a_g (x_0, z_0) + I_{gh}(0, y_0) = 0$.     
In fact, we regularize the gluon contribution to the vacuum energy density  
by substracting unknown term by means of another unknown (arbitrary) ghost 
contribution, i. e. $ \epsilon_g + \epsilon_{gh} = \epsilon^{reg}_g = (\pi x_0)^{-2} p_0^4 I_g (0, x_0)$ and in what follows superscript "reg" will be omitted
for simplicity. It is easy to show that the
above mentioned condition of cancellation $\Delta =0$ of the UV divergences (due to arbitrary $z_0, \ y_0$) is consistent with this definition of the regularized vacuum energy density of the nonperturbative gluon part. At the same time, 
it becomes negative and has a local minimum (see below) while the unregularized
contribution (depending on $x_0$) $(\pi x_0)^{-2} p_0^4 I^a_g (0, x_0)$  is always positive (because of constant $a$, which precisely violates normalization condition). Thus our regularization procedure is in agreement with general physical interpretation of ghosts to cancel the effects of the unphysical degrees of freedom of the gauge bosons [31,32].
    
Thus the regularized vacuum energy density due to the  nonperturbative gluon contributions becomes

\begin{equation}
\epsilon_g =  {1 \over \pi^2} p_0^4 x_0^{-2} \times 
\Bigl\{ {9 \over 2} \ln(1 + {x_0 \over 3}) - {3 \over 4} x_0
- {1 \over 2} x^2_0 \ln(1 + {3 \over x_0}) \Bigr\}.
\end{equation}
Let us introduce the effective potential at fixed $p_0$, namely

\begin{equation}
\Omega_g = { 1 \over p_0^4} \epsilon_g.                      
\end{equation}
Its behaviour as a function of $x_0$ is shown in Fig. 1. It has a few remarkable features. First, it has the local minimum at $x^{min}_0 = 4 \ln(1 + {x^{min}_0 \over 3}) = 2.2$ (stationary condition) 
and zero at $x_0^0 = 0.725 $. Second it asymptotically vanishes at $x_0 \rightarrow \infty$. This means that it satisfies the normalization condition since at
fixed $p_0$ the parameter $x_0$ can go to infinity when $\bar \mu^2$ goes to zero (recalling the definition  $x_0 = p_0^2 / \bar \mu^2$), so indicating that in this limit there is no the nonperturbative phase at all and only the perturbative phase remains. However, it still has a defect since the opposite limit $x_0 \rightarrow 0$ ($\bar \mu^2 \rightarrow \infty$) is unphysical. The problem is that the confinement scale is zero or finite, it 
can not be arbitrarily large. That is why the vacuum energy density in this limit
becomes positive (see Fig. 1). This obviously means that the physical region for parameter $x_0$ is bounded from below, namely $x_0 \ge x_0^0 = 0.725$. In this region the vacuum energy density is always negative as it should be.             
At the same time the value of the vacuum energy density at the stationary state
does not depend on this unphysical tail, of course (see next section). Thus it 
has no imaginary part (our vacuum is stable) and at stationary state it is

\begin{equation}
\epsilon_g = \epsilon_g (x_0^{min}, p_0) = - 0.0263 p_0^4,
\end{equation}
where $p_0$ determines the scale of nonperturbative dynamics within our approach. If QCD confines such characteristics scale should certainly exist.

\section{ The gluon condensate and topological susceptibility}

The vacuum energy density
is important in its own right as the main characteristics of the
nonperturbative vacuum of QCD. Futhermore it assists in
estimating such an important phenomenogical parameter as
the gluon condensate, introduced in the QCD sum rules approach
to resonance physics [33]. 
The famous trace anomaly relation [34] in the general case (nonzero current quark masses $m_f^0$) is             

\begin{equation}
\Theta_{\mu\mu} = {\beta(\alpha_s) \over 4 \alpha_s}
G^a_{\mu\nu} G^a_{\mu\nu}
+ \sum_f m_f^0 \overline q_f q_f.
\end{equation}
where $\Theta_{\mu\mu}$ is the trace of the energy-momentum
tensor and $G^a_{\mu\nu}$ being the gluon field strength tensor.
Sandwiching (3.1) between vacuum states and on account of the obvious relation 
$\langle{0} | \Theta_{\mu\mu} | {0}\rangle = 4 \epsilon_t$, one obtains

\begin{equation}
\epsilon_t = {1 \over 4} 
\langle{0} | {\beta (\alpha_s)  \over 4 \alpha_s} G^a_{\mu\nu} G^a_{\mu\nu} | {0}\rangle
+ {1 \over 4} \sum_f m^0_f \langle{0} | \overline q_f q_f | {0}\rangle,
\end{equation}
where $\epsilon_t$ is the sum of all possible contributions
to the vacuum energy density and 
$\langle{0} | \overline q_f q_f | {0}\rangle$ is the quark condensate. From this equation in the chiral limit ($m^0_f=0$) and saturating $\epsilon_t$ by $\epsilon_g$, one obtains

\begin{equation}
\langle \bar G^2 \rangle = -
\langle{0} | {\beta(\alpha_s) \over 4 \alpha_s} G^a_{\mu\nu} G^a_{\mu\nu} | {0}\rangle = - 4 \epsilon_t = - 4 \epsilon_g,
\end{equation}
where $\epsilon_g$ is given by (2.11). Let us note that if confinement happens 
then the $\beta$ function is always in the domain of attraction   
(i. e. always negative) without IR stable fixed point [31]. Thus the nonperturbative gluon condensate $\langle \bar G^2 \rangle$ in the strong coupling limit, defined in (3.3), is always positive as it should be. 

One of the main characteristic of the QCD vacuum is the topological density    
operator (topological susceptibility) in gluodynamics

\begin{equation}
\chi_t =  \lim_{q \rightarrow 0} i \int d^4x\, e^{iqx} \langle{0} | T \Bigl\{ {q(x) q(0)} \Bigr\} | {0} \rangle,   
\end{equation}
where $q(x)$ is the topological charge density, defined as
$q(x) =  {\alpha_s \over 8 \pi} G (x) \tilde{G} (x) = {\alpha_s \over 8 \pi}
G^a_{\mu \nu} (x) \tilde {G}^a_{\mu \nu}(x)$    
and $\tilde {G}^a_{\mu \nu}(x) = {1 \over 2}
\epsilon^{\mu \nu \rho \sigma} G^a_{\rho \sigma} (x)$
is the dual gluon field strength tensor. In definition of the topological 
susceptibility (3.4) it is assumed that the corresponding regularization and subtraction have been already done in order (3.4) to stand for the renormalized and finite topological susceptibility (see Refs. [35-37]). 

As it was shown in Ref. [36], the topological susceptibility can be related to 
the nonperturbative gluon condensate via the low energy theorem in gluodynamics
[35,36] as follows 

\begin{equation}
\lim_{q \rightarrow 0} i \int d^4x\, e^{iqx} \langle{0} | T \Bigl\{ {\alpha_s \over 8 \pi} G \tilde{G} (x)     
{\alpha_s \over 8 \pi} G \tilde {G}(0) \Bigr\} | {0} \rangle =  \xi^2 \langle { \beta(\alpha_s) \over 4 \alpha_s} G^2 \rangle. 
\end{equation}
There exist two proposals to fix the numerical value of the coefficient 
$\xi$. The value $\xi = 2/b, \ b=11$ was suggested long time ago by Novikov, Schifman, Vanshtein and Zakharov (NSVZ), who used the dominance of self-dual fields hypothesis in the YM vacuum [35].                 
A second one, $\xi = 4/3b$, was advocated very recently by Halperin and Zhnitsky (HZ), using a one-loop connection between the conformal and axial anomalies in the theory with auxiliary heavy fermions [36]. It has been confirmed
by completely different method [37]. However, in our numerical calculations we 
will use both values for the $\xi$ parameter. We    
note that there exists an interesting relation between the HZ and NSVZ values, 
namely $\xi_{HZ} = (2/N_c)\xi_{NSVZ}, \ N_c=3$. Using our values for the nonperturbative gluon condensate (3.3), the topological susceptibility (3.4) can be easily calculated 
      
\begin{equation}
\chi_t = - \xi^2 \langle { \beta(\alpha_s) \over 4 \alpha_s} G^2 \rangle  
= - (2 \xi)^2 \epsilon_g.
\end{equation}
Let not note that in the derivation of (3.6) from (3.5) we replaced 
$ \beta(\alpha_s) \rightarrow - \beta(\alpha_s)$ in accordance with our definition of the gluon condensate in the strong coupling limit, (3.3).
 The topological susceptibility (3.6)
asists in the resolution of the $U(1)$ problem (the large mass of the $\eta'$ meson) within the Witten-Veneziano (WV) formula for the mass of the $\eta'$ meson [38].  Within our notations it is expressed as follows 
$f^2_{\eta'} m^2_{\eta'} = (16 N_f/ N_c^2) \chi_t$,
where $f_{\eta'}$ is the $\eta'$ residue defined as 
$\langle {0}| \sum_{q=u,d,s} \overline q \gamma_{\mu} \gamma_5 q | {\eta'} \rangle = i \sqrt{N_f} f_{\eta'} p_{\mu}$ and
$\langle {0}| N_f {\alpha_s \over 4 \pi} G \tilde{G} | {\eta'} \rangle = (N_c \sqrt{N_f} / 2) f_{\eta'} m^2_{\eta'}$. So, following Witten [38], the anomaly equation is
$\partial_{\mu} J_5^{\mu} = 2 N_f ({2 \over N_c}) {\alpha_s \over 8 \pi} G \tilde{G}$.
Here and below $N_c=3$ is the number of colors. Using also the normalization relation $f_{\eta'} = \sqrt{2} F^0_{\pi}$, one finally obtains
             
\begin{equation}
F^2_{\pi}m^2_{\eta'} = {8 N_f \over N_c^2} \chi_t = 2 N_f ({2 \over N_c})^2 \chi_t = 2 N_f \chi_t^{WV}
\end{equation}
where for simplicity we omit the superscript "0" in the pion decay constant as 
well as in $m^2_{\eta'}$ and $\chi_t$. In the numerical evaluation of the above formula, we will put, of course, $N_f=N_c=3$, while the topological susceptibility will be evaluated at $N_f=0$ as it should be by definition.     

As was mentioned above, there exists also a contribution to the total vacuum energy density at the classical level given by the instanton-type nonperturbative
fluctuations of gluon fields. Within RILM [5] it is given by     

\begin{equation}
\epsilon_I = - {b \over 4} n = - {b \over 4} \times 1.0 \ fm^{-4}       
= - (0.00417 - 0.00025N_f ) \ GeV^4,
\end{equation}
where $b$ is the first coefficient of the $\beta$-function (see below) and 
$n$ is the density of the instanton-type fluctuations in the QCD vacuum.
The CS-GML function $\beta(\alpha_s)$, up to terms of order $\alpha^3_s$, i. e.
in the weak coupling limit, is:

\begin{equation}
\beta(\alpha_s) = - b { \alpha^2_s \over 2 \pi} + O(\alpha^3_s), \qquad        
 b = 11 - {2 \over 3} N_f.
\end{equation}

Recently, in quenched ($N_f=0$) lattice QCD by using the so-called "cooling" method the role of the instanton-type fluctuations in the QCD vacuum was investigated. In particular, Negele et al [39] found that the instanton density should be $n= (1+ \delta) \ fm^{-4}$, where $\delta \simeq 0.3-0.6$ depending on cooling steps. Moreover, by studying the topological content of the vacuum of $SU(2)$ pure gauge theory using also lattice simulations [11], it is concluded that the average radius of an instanton is about $0.2 \ fm$, at a density
of about $2 \ fm^{-4}$. Thus at this stage it is rather difficult to choose    
some well-justified numerical value of the instanton-type contribution to the total vacuum energy density.                                     

It becomes almost obvious that we must  distinguish between the two types of gluon condensates, nevertheless, both of which are nonperturbative quantities. The first
one is determined by (3.3)
and it is the one which is relevant in the strong coupling limit. In this case
the total vacuum energy is mainly saturated by the ZME component as it 
is precisely shown in (3.3). In the weak coupling limit, one may use the solution (3.9) for the $\beta$-function. Then Eq. (3.3) becomes 

\begin{equation}
\langle G^2 \rangle =
 \langle {\alpha_s \over \pi} G^2 \rangle \equiv
\langle{0} | {\alpha_s \over  \pi} G^a_{\mu\nu} G^a_{\mu\nu} | {0}\rangle =    
- {32 \over b} \epsilon_t = - {32 \over b} \epsilon_I = 8 n,
\end{equation}
and now the total vacuum energy density is mainly saturated by $\epsilon_I$, Eq. (3.8), as it is already shown in Eq. (3.10), since $\epsilon_g$ becomes very small in the weak coupling limit (see equation (2.11) at $p_0 \rightarrow 0$). Precisely this gluon condensate was introduced long ago [33].  The topological susceptibility due to instantons is again determined by (3.6) by substitution $\epsilon_g \longrightarrow \epsilon_I$, where
$\epsilon_I$ is given (3.8) in the case of RILM. In order to obtain 
Negele et.al [39] and DeGrand et.al [11] values, the RILM value should be simply multiplied by factors (1.3-1.6) and 2, respectively. 
However, we will not explicitly calculate the DeGrand et.al mode since it stands for pure $SU(2)$ gauge theory.  

Let us note that saturating the right hand side of (3.10) by the
instanton component (3.8), the gluon condensate becomes not explicitly dependent on number of flavors $N_f$.                                          
This unphysical situation takes place because in instanton calculus [5-7] there
is no other way to calculate the vacuum energy density than the 
trace anomaly ralation ((3.2-3.3) which become finally (3.10) by substitutinginto the left hand side of (3.10) phenomenologically extracted value of the
gluon condensate. In this case it is preferable to have the $N_f$ dependent va 
cuum energy density than the gluon condensate since the former is the main cha racteristic of the nonperturbative vacuum. Contrast to this, we have calculated
the vacuum energy density completely independly from the trace anomaly relation. We use it only to calculate the gluon condensate in the strong and weak coupling limits. That is why in our case both quantities are, in principle, $N_f$ dependent functions though in the present work we are restricted by the quenched 
QCD.   

In conclusion, let us make a few things perfectly clear. It makes sense to underline once more that the vacuum energy density $is \ not$ determined by the trace anomaly relation (3.2). The real situation is completely opposite. As was mentioned in Introduction, the nonperturbative QCD vacuum contains many different
components and ingredients which contribute to the vacuum energy density. These
contributions are completely independent from the gluon condensate, of course. 
Thus, the total vacuum energy density, 
defined as the trace of the energy-momenum tensor, becomes the sum of all independent contributions. This sum precisely determines the realistic value of the 
gluon condensate in the chiral limit via the trace anomaly ralation. Here we establish the quantum part of the total vacuum energy density which is due to $q^{-4}$ nonperturbative, topologically nontrivial
excitations of the gluon field configurations in the QCD vacuum. It is given by
the effective potential for composite operators in the form of loop expansion  
where the number of 2PI vacuum loops (consisting of confining quarks and nonperturbative $q^{-4}$ gluons, properly regularized with the help of ghosts) is equal to the power of the Plank constant. So the quantum part of the vacuum energy
density to leading order becomes $\epsilon_g + N_f \epsilon_q +O(h^2)$ and, in 
principle, cannot be accounted for by the trace anomaly relation. The instanton-type topological fluctuations, being a classical phenomena, nevertheless also contribute to the vacuum energy density through a tunneling effect which is known to lower the energy of the ground state [5]. In the present work we will simply attempt to sum up these two (quantum and classical) well-established contributions to the vacuum energy density.

\section{Numerical results}

It is instructive to start analysis of our numerical investigation of the true QCD topology by reproducing  the WV formula (3.7) in the nonchiral case as well, namely
  
\begin{equation}
m^2_{\eta'} = {2 N_f \over F^2_{\pi}} \chi_t^{WV} + \Delta ,
\end{equation}
where $\Delta = 2 m^2_{K} - m^2_{\eta}$. In what follows all numerical results for the topological susceptibility stand for $\chi_t^{WV}$ and not for $\chi_t$, shown in (3.7). However the superscript "WV" will be omitted for convenience.
The precise validity of the WV formula (4.1) is, of course, not completely clear,
nevertheless, let us regard it as exact (for simplicity). Using now experimental values of all physical quantities
entering this formula, one obtains that the phenomenological ("experimental")  
 value of the topological susceptibility is 

\begin{equation}
\chi_t^{phen} = 0.001058 \ GeV^4 = (180.36 \ MeV )^4.
\end{equation}
In the chiral limit $\Delta = 0$ since $K^{\pm}$ and $\eta$ particles are NG bosons. Omitting formally this contribution from the right hand side of Eq. (4.1 ), we are able to derive an upper bound for the mass of the $\eta'$ meson in the chiral limit

\begin{equation}
m^0_{\eta'} \lesim 0.8542 \ GeV = 854.2 \ MeV,
\end{equation}
which is comparable with its experimental value $m^{exp}_{\eta'}  = 957.77 \ MeV$.

 The main problem now how to set the scale at which our calculations should be 
done. To set the scale within our approach means to choose a reasonable value for the scale parameter $p_0$ in (2.12). Only the requirement is that the scale-setting scheme should be physically well-motiviated since $p_0$ determines the 
scale of the nonperturbative dynamics in our approach, so it can not be arbitrary large. We beleive that $p_0=1 \ GeV$ is a realistic upper bound for the scale responsible for the nonperturbative dynamics in chiral QCD. Our numerical investigation of the chiral QCD topology stracture at this scale is presented in Ref. [40]. Let us underline 
that its value has nothing to do with the values chosen to analyse numerical results in phenomenology (for example, $1 \ GeV$ or $2 \ GeV$) which have no physical sense and are simply convention, while our scale has a direct and clear physical meaning separating nonperturbative phase from the perturbative one. However, here we intend to use more sophisticated scale-setting schemes in order to
numerically determine it.                                                      
          
In our previous publications [14,41] the expression for the pion decay constant
was derived

\begin{equation}
F^2_{CA} = {3\over {8 \pi^2}}p_0^2z_0^{-1}
             \int^{z_0}_0 dz \,{ zB^2(z_0, z) \over
             {\{zA^2(z) + B^2(z_0, z)\}}}.
\end{equation}
as well as for the dynamically generated quark mass which was defined as the 
inverse of the quark propagator at zero point

\begin{equation}
\Lambda_{CSBq} = 2 m_d = 2 p_0 \bigl\{z_0 B^2(z_0,0)\bigr\}^{-1/2},
\end{equation}
where $\Lambda_{CSBq}$ defines the scale at which DCSB occures at the fundamental quark level (see also Ref. [42]).                                          
The pion decay constant in general case (chiral or non-chiral) does not depend 
on how many different quark flavors occur in the QCD vacuum, while, for example, the value of the quark condensate does depend on it. On the other hand, the constant of integration $z_0$ entering these expressions need
not be numerically the same as in the evaluation of the vacuum energy density in section 2 though the definition is, of course, the same. That is why we denote here it as $z_0$ (not to be mixed up with $z_0$ from (2.6-2.7)). Its numerical value may depend, in principle, on normalization condition, boundary conditions, etc.                                                                    
At the same time, the mass scale parameter $p_0$ as determining the scale of nonperturbative dynamics in general is consedering here as unique. In other words
it is understood that this scale is commom for the chiral QCD topology 
as well as for the chiral physical parameters such as the pion decay constant, 
dynamically generated quark mass, ets.  

In order to determine the mass scale parameter $p_0$, 
which characterizes the region where confinement, DCSB and
other nonperturbative effects are dominant, we propose
to use the following bounds for the pion decay constant in the chiral limit
 
\begin{equation}
87.2 \leq  F^o_{\pi} \leq 93.3  \ (MeV).
\end{equation} 
The pion decay constant in the chiral limit, evidently, can not exeed its experimental value, so the upper bound in (4.6) is uniquely well-fixed. The lower
bound in (4.6) is fixed from 
the chiral value of the pion decay constant as obtained in Ref. [43], namely $F^o_{\pi} = (88.3 \pm 1.1) \ MeV$, which, obviously, satisfies
(4.6). The value $F_{\pi} = 92.42 \ MeV$, advocated in Refs. [44], also satisfies these bounds. We think that chosen interval covers all the realistic values
of the chiral pion decay constant. In any case it is always possible to change 
lower bound to cover any requested value of the pion decay constant in the chiral limit. This bound is chosen as unique input data in our numerical investigation of chiral QCD.
The pion decay constant is a good experimental number since it is a directly
measurable quantity in contrast, for example, to the quark condensate or dynamically generated quark mass.
For this reason our choice (4.6) as input data opens up
the possibility of realibly estimating the deviation of the chiral
values from their "experimental" (empirical) phenomenologically determined
values of various physical quantities which can not be directly measured.      
Thus to assign definite values to the physical
quantities in the chiral limit is a rather delicate question (that is why we prefer to use and obtain bounds for them rather than the definite values). At
the same time it is a very important theoretical
limit which determines the dynamical structure of low-energy QCD.
We set the scale by two different way, which nevertheless lead to almost the sme numerical results.

\subsection{Analysis of the numerical data at the minimium of the effective    
            potential}

It is quite resonable to set the scale by identifying the constant of integration $z_0$ in (4.4-4.5) with the value of the constant of integration $x_0$ at which the minimum of the effective potential (2.11) occures, namely $x_0^{min}=2.2$ (see Fig. 1). However, as was noticed at the end of
section 2, the location of this minimum is affected by the contributions from unphysical region of small $x_0 \le x_0^0 = 0.725$. That is why we can not directly identify $z_0$ with the above mentioned minimum. The resolution of this problem is
rather simple. Let us introduce a new variable in the effective potential (2.10-2.11) as follows $z_0 = x_0 - x_0^0$, which describes a constant shift only. 
Then the position of the minimum of the effective potential (as a function of $z_0$) is given now by the following stationary condition

\begin{equation}
z^{min}_0 = - x_0^0 + 4 \ln(1 + {z^{min}_0 + x_0^0 \over 3}) = 1.475
\end{equation}  
and it is not affected by unphysical contributions since now the effective 
potential as a function of $z_0$ is always negative (see Fig. 2). At the same time the value of the vacuum energy density at the stationary state remains, of
course, unchanged by this constant shift, i. e. 

\begin{equation}
\epsilon_g = \epsilon_g (x_0^{min}, p_0) = \epsilon_g (z_0^{min}, p_0) = - 0.0263 p_0^4.
\end{equation}
Identifying now simply

\begin{equation}
z_0 = z^{min}_0 = 1.475,
\end{equation}
and again using (4.6), one obtains                                             

\begin{equation}
680.76 \leq p_0 \leq 728.4  \ (MeV)
\end{equation}   
and

\begin{eqnarray}
292.8 &\leq m_d \leq& 313.3  \ (MeV), \nonumber\\
560.45 &\leq \bar \mu \leq& 600  \ (MeV)
\end{eqnarray}
We are able now to numerically evaluate the chiral QCD topological structure. Results of our calculations are shown in Table 1 (calculation scheme A).

\subsection{Analysis of the numerical data at a
             scale of DCSB at the quark level}

  There is a natural scale in our approach to DCSB. At the
fundamental quark level the chiral symmetry is spontaneously broken
at a scale $\Lambda_{CSBq}$ defined in (4.5). We may then
analyse our numerical data at a scale at which DCSB at the
fundamental quark level occurs. For this aim, what is needed is
only to simply identify mass scale parameter $p_0$
with this scale $\Lambda_{CSBq}$, i.e. to put

\begin{equation}
p_0 \equiv \Lambda_{CSBq} = 2 m_d.
\end{equation}
Now one can uniquely determine the constant of integration of the quark SD equation. Indeed, from (4.5) it immediately follows that this constant is equal to

\begin{equation}
z_0 = 1.34805.
\end{equation}
Using the bounds for the pion decay constant (4.6), one obtains
                
\begin{equation}
715.24 \leq p_0 \leq 765.28  \ (MeV)
\end{equation}
and

\begin{eqnarray}   
357.62 &\leq m_d \leq& 382.64  \ (MeV), \nonumber\\                            
616 &\leq \bar \mu \leq& 659.13  \ (MeV)                                     
\end{eqnarray}
on account of (4.7). This means that 
all physical quantities considered in our paper are uniquely determined. The results of our calculations are displayed in Table 2 (calculation scheme B).
Let us only note in advance that (4.13) is remarkably close to (4.9) which was 
determined from completely different source.

\section{Summary}

That the QCD vacuum has a stationary state is a result of the existence and importance of the nonperturbative $q^{-4}$ quantum excitations of the gluon field 
configurations there. The location of the minimum at which stationary state occurs, on the other hand, determines the constant of integration of the corresponding quark SD equation (calculation scheme A). The value of the scale responsible for the nonperturbative dynamics, however, is taken from the bounds for the 
pion decay constant in the chiral limit (4.6) which is unique input data within
our approach. We have obtained rather resonable values for this scale (4.10) as
well as for the values of the dynamically generated quark mass, $m_d$ (4.11).
It seems to us that the calculation scheme B sligthly overestimates these 
values, especially for $m_d$ (see (4.14-4.15)). Just recently the effects of nonperturbative QCD in the 
nucleon structure functions was discussed. A universal mass scale parameter 
$m_a \simeq 470 \ MeV$ of the nonperturbative dynamics in QCD was obtained [45]. This is in rather good agreement with our numerical results for the mass scale parameter $\bar \mu$ (4.11) taking into account completely different physical observables have been analysed. 

Our numerical results for the chiral QCD topology are presented in Tables 1 and
2. In general our values for the vacuum energy density approximately two times 
bigger than RILM's value and comparabale with values due to Negele et.al. The  
topological susceptibility in the chiral limit can not, of course, exeed
its phenomenological ("experimental") value given in (4.2). Thus all our numerical results for this quantity are in fair agreement with it especially those of
the NSVZ value for the $\xi$ parameter, introduced in the low-energy theorem, Eq. (3.5), while HZ mode substantially underestimates    
phenomenological value of the topological susceptibility as well as the value of the mass of the $\eta'$ meson in the chiral limit, Eq. (4.3).                
Instanton contributions also substantially underestimate the experimental value
of the topological susceptibility and therefore can not account for the large mass of the $\eta'$ meson (see Table 3). They also are in disagreement with recent lattice results. The topological susceptibility   
in pure $SU(3)$ gauge theory was determined in Ref. [46] by using an improved topological charge density operator and it is $\chi_t^{1/4} = 175 (5) \ MeV$.  
The topological properties of $SU(3)$ gauge theory using improved cooling method was studied in Ref. [47] and it is $\chi_t^{1/4} = 182 (8) \ MeV$. Obviously,
our results are consistent with these values. Very promising new lattice  
smoothing process based on the renormalization group equation which removes
short distance fluctuations but preserves long distance structure was proposed 
in Ref. [11]. The topological content of the vacuum of still yet 
$SU(2)$ pure gauge theory was studied there. 
 
 As it was mentioned in section 3, the vacuum energy density is, in principle, 
the sum of all possible independent contributions, at least is the sum of two well-defined contributions, quantum $\epsilon_{ZME}= \epsilon_g$ and classical $\epsilon_I$, i. e. $\epsilon_t = \epsilon_g + \epsilon_I + ...$,
where the dotts denote other possible independent contributions. In this case  
indeed an excellent agreement with phenomenology is achived (see Tables 4 and 5). In any way, it is clear that the instanton-type fluctuations provide dominant nonperturbative component contributing to the total vacuum energy only in the
weak coupling limit. The dominant nonperturbative contribution to the total 
vacuum energy in strong coupling limit is provided by the ZME
model of the QCD vacuum. At large distances instantons remain one of possible contributions while at short distances they become apparently the dominant component since the ZME contributions at these distances quickly vanish (see (4.8) at $p_0 \rightarrow 0$).

The phenomenological analysis of QCD sum rules [33] for the gluon condensate implies

\begin{equation}
\langle{0}|{\alpha_s \over
\pi}G^a_{\mu\nu}G^a_{\mu\nu}|{0}\rangle \simeq  0.012 \ GeV^4,
\end{equation}
This estimate can be changed within a factor of two [33]. However, it was ponted out (perhaps first) in Ref. [48] (see also Ref. [49]) that QCD sum rules substantially underestimate the value of the gluon condensate. The numerical value 
of the gluon condensate in RILM of the QCD vacuum, for a dilute ensemble is shown in Table 3. Of course it completely coincedes with QCD sum rules value (5.1)
since in RILM the parameters
characterizing the vacuum, the instanton size $\rho_0 = 1/3 \ fm$
and the "average separation" $R= 1.0 \ fm$ were chosen
to precisely reproduce traditional (phenomenologically estimated
from QCD sum rules) values of quark and gluon condensates (5.1), respectively.
However, the most recent phenomenological calculation of the gluon condensate is given by Narison in Ref. [50], where a brief review of many previous calculations is also presented. His analysis leads to the update average value as

\begin{equation}
\langle{0}|{\alpha_s \over
\pi}G^a_{\mu\nu}G^a_{\mu\nu}|{0}\rangle = (0.0226 \pm 0.0029) \ GeV^4.
\end{equation} 
This value is in good agreement with our results for the gluon condensate in the weak coupling limit (obtained with the help of (3.10) when $\epsilon_t$ is saturated by $\epsilon_g$) and when the pion decay constant in the chiral limit is approximated by its experimental value (see Tables 1 and 2). 

In conclusion one remark is in order. We have taken into account the instanton-induced interaction purely phenomenologically by simply summig up the two contributions (classical instanton's and quantum ZME) to the vacuum energy density. 
How to take into account the instanton-induced interaction at the fundamental quark level within our approach is not completely clear for us though see paper 
[51] and references therein.

\acknowledgements

  The author is grateful to the late Prof. V.N. Gribov for many
useful remarks and discussions on nonperturbative QCD. He also would like to thank I. Halperin for correspondence. He also would like to thank H.Suganuma and 
espicially H.Toki and T.Sakai for useful remarks, discussions and help in numerical calculations. This work was supported by Special COE grant of the Ministry
of Education, Sience and Culture of Japan.

\vfill

\eject

 \vfill

 \eject

\begin{figure}
\caption{The effective potential (2.10-2.11) as a function of $x_0$. For       
         details see corresponding places in the main body of the text.}

\caption{The effective potential (2.10-2.11) as a function of $z_0$. For       
         details see corresponding places in the main body of the text.
}
\end{figure}

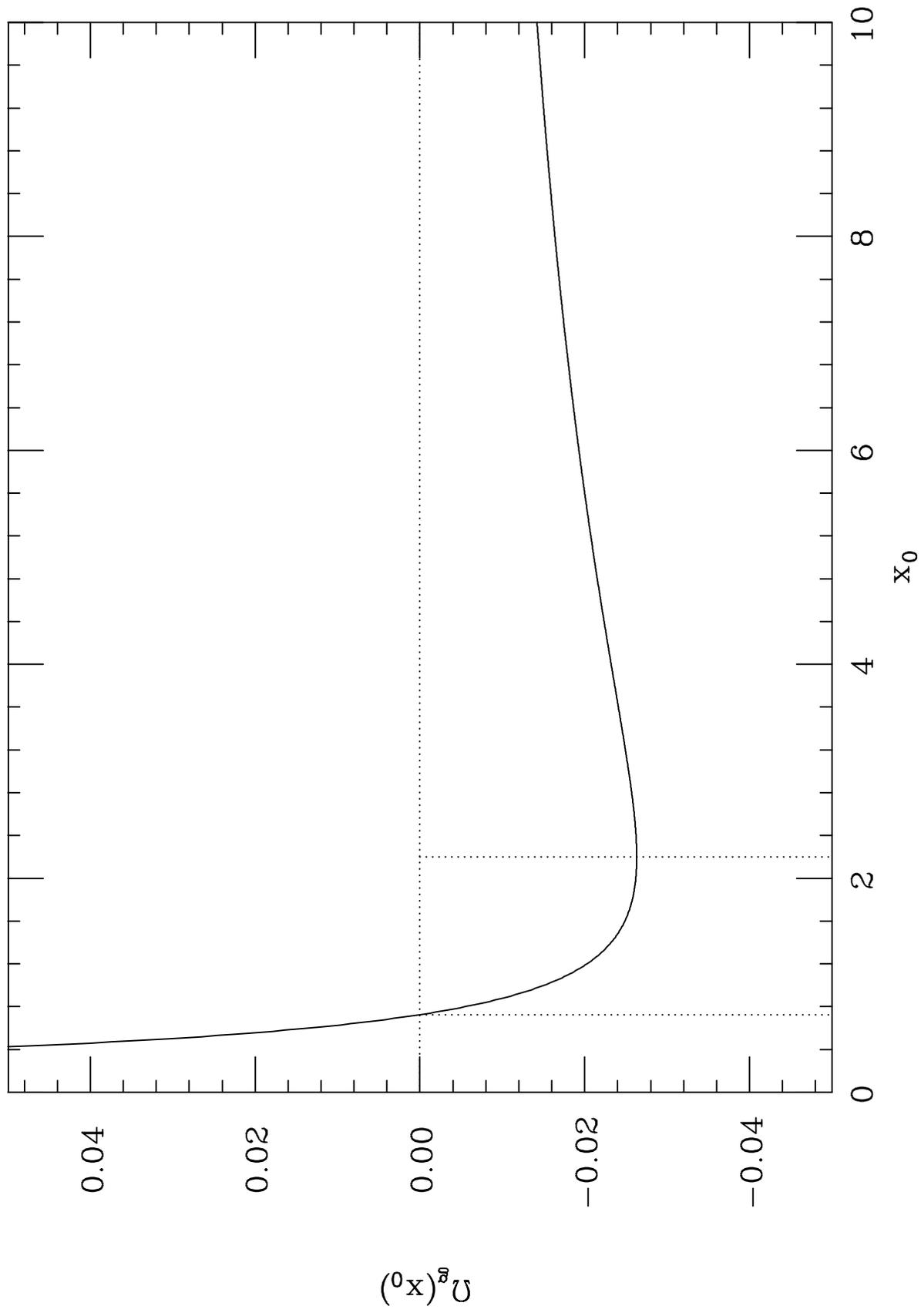
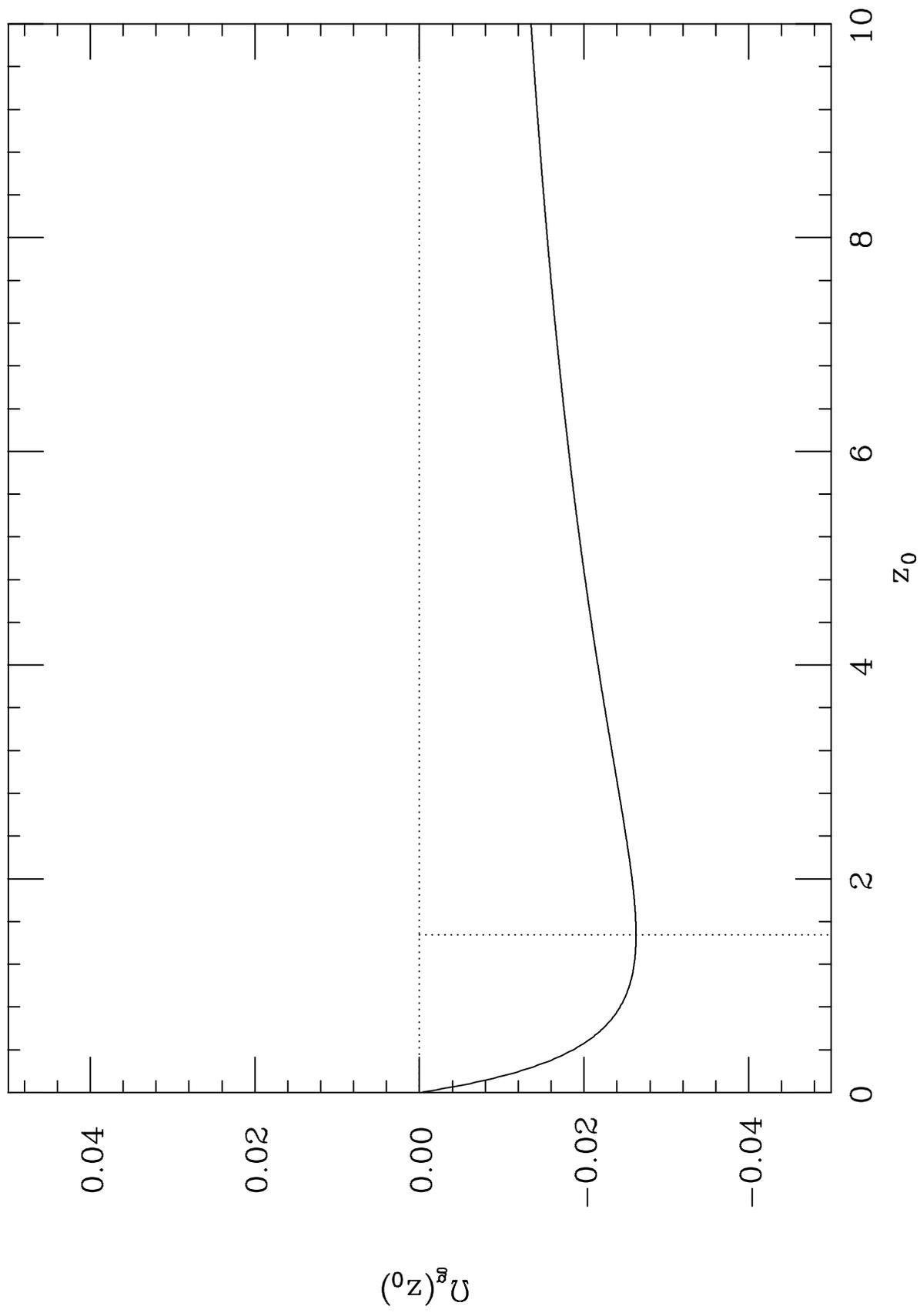
\end{document}